\def\be{\begin{equation}}
\def\ee{\end{equation}}
\def\ba{\begin{array}}
\def\ea{\end{array}}

\documentclass[preprint,showpacs,amsmath]{revtex4}
\usepackage{amsfonts}
\usepackage{amssymb}
\def\qed{\leavevmode\unskip\penalty9999 \hbox{}\nobreak\hfill
     \quad\hbox{\leavevmode  \hbox to.77778em{%
               \hfil\vrule   \vbox to.675em%
               {\hrule width.6em\vfil\hrule}\vrule\hfil}}
     \par\vskip3pt}

\input amssym.def

\begin{document}
\parskip=4pt
\parindent=18pt

\parskip=4pt
\parindent=18pt
\baselineskip=22pt \setcounter{page}{1} \centerline{\Large\bf
Separable State Decompositions for a Class of Mixed States}
\vspace{6ex}

\begin{center}
 Ting-Gui Zhang$^1$, Xiaofen Huang$^{*,2}$,Xianqing Li-Jost$^1$, Naihuan
 Jing$^{3,4}$ and Shao-Ming Fei$^{1,5}$
\bigskip

\begin{minipage}{6.5in}
$^1$ Max-Planck-Institute for Mathematics in the Sciences, Leipzig, 04103, Germany\\
$^2$ School of Mathematics and Statistics, Hainan Normal University, Haikou, 571158, China\\
$^3$ School of Sciences, South China University of
Technology, Guangzhou, 510640,  China\\
$^4$ Department of Mathematics, North Carolina State University,
Raleigh, NC27695, USA\\
$^5$ School of Mathematical Sciences, Capital Normal University, Beijing
100048, China\\
$*$ Corresponding author, e-mail address: huangxf1206@googlemail.com
\end{minipage}
\end{center}
\vskip 1 true cm
\begin{center}
\begin{minipage}{5in}
\vspace{3ex} \centerline{\large Abstract} \vspace{4ex}

We study certain quantum states for which the PPT criterion is both
sufficient and necessary for separability. A class of $n\times n$
bipartite mixed states is presented and the conditions of PPT for
these states are derived. The separable pure state decompositions of
these states are explicitly constructed when they are PPT.
\end{minipage}
\end{center}
\bigskip
\medskip

Quantum entangled states have become one of the key resources in
quantum information processing. The study of quantum teleportation,
quantum cryptography, quantum dense coding, quantum error correction
and parallel computation \cite{1,2,3} has spurred a flurry of
activities in the investigation of quantum entanglement. Despite the
potential applications of quantum entangled states, there are many
open questions in the theory of quantum entanglement. The
separability of quantum mixed states is one of the important
problems in the theory of quantum entanglement.

Let $H$ be an $n$-dimensional complex Hilbert space, with
$|i\rangle$, $i=1, ..., n$ the orthonormal basis. A bipartite mixed
state in $H\otimes H$ is said to be separable if the density matrix
can be written as
\begin{equation}\label{sp}
\rho=\sum_i p_i\rho_i^1 \otimes \rho_i^2,
\end{equation}
where $0<p_i\leq 1$, $\sum_i p_i=1$, $\rho_i^1$ and $\rho_i^2$ are
density matrices associated with the first and the second Hilbert
spaces respectively. It is a challenge to find a decomposition like
(1) or to prove that such a decomposition does not exist for a given
state $\rho$. With considerable effort in analyzing the
separability,  there have been some (necessary) criterias for
separability in recent years, for instance, Bell inequalities
\cite{bell}, PPT (positive partial transposition) \cite{peres},
reduction criterion \cite{mp,cerf}, majorization criterion
\cite{nielsen}, entanglement witnesses \cite{terhal,lewenstein},
realignment \cite{rudolph,chen2} and generalized realignment
\cite{fei1}, range criterion \cite{l31}, criteria based on the local
uncertainty relations \cite{l434546}, correlation matrix approach
\cite{l474849}, as well as some necessary and sufficient criterias
for low rank density matrices \cite{pm,fei2,fei3}.

The PPT criterion is generally a necessary condition for
separability. It becomes also sufficient for the cases of $2\otimes 2$ and
$2\otimes 3$ bipartite states \cite{mpr2}. In \cite{vidal}, it has
been shown that a state $ \rho$ supported on $ m\times n$
Hilbert space ($m\leq n$) with $rank( \rho)\leq m$ is separable if
and only if $ \rho$ is PPT. However, it is generally a difficult problem to
find the concrete PPT conditions for a given such state within this class.
Moreover, even if the PPT conditions are satisfied and hence the state is separable,
it is still a challenging problem to find the detailed separable pure state decompositions (\ref{sp}).
For separable two-qubit states, an elegant separable pure state decompositions has been given in \cite{Wootters98}.

In \cite{feili} a class of $3\otimes3$ mixed states $\rho$ with
$rank( \rho)= 3$ has been investigated. The PPT conditions are
derived. And the explicit separable pure state decompositions are
constructed. In this paper we generalize the results in \cite{feili}
to a class of $n\otimes n$ quantum mixed states. We derive the PPT
conditions and construct explicitly the separable pure state
decompositions for states satisfying the PPT conditions.

We consider a set of mixed states defined in $H\otimes H$ space
which has the following form of spectral decomposition:
\begin{equation}\label{state}
\rho=\sum_{l=1}^{n}\lambda_l| V_l\rangle\langle V_l |,
\end{equation}
with $\sum_{l=1}^{n}\lambda_l=1$, $0< \lambda_l<1 $, and
\be\label{7} |V_l\rangle=\sum_{j=1}^{n}
v_l^{j}|j\rangle\otimes|j+l-1\rangle,~~ l=1,2,\cdots,n , \ee where
$0\neq v_l^{j}\in \Bbb{C}$ and $\sum_j \overline{v_l}^{j}v_l^{j}=1$.
($\overline{z}$ denoting the complex conjugation of $z$). When
$n=3$, the state $\rho$ becomes the object of study in \cite{feili}.
For simplicity we denote $x_l^{j}=\sqrt{\lambda_l}v_l^j$,
$|X_l\rangle=\sqrt{\lambda_l}|V_l\rangle$. Then $\rho$ has the form,
$$
\rho=\sum_{l=1}^n|X_l\rangle\langle
X_l|=\sum_{l,j,k=1}^nx_{l}^{j}\overline{x}_{l}^{k}|j\rangle\otimes
|j+l-1\rangle\langle k|\otimes
\langle k+l-1|=\sum_{l,j,k=1}^nx_{l}^{j}\overline{x}_{l}^{k}|j\rangle\langle
k|\otimes|j+l-1\rangle\langle k+l-1|.
$$
We first deduce the PPT conditions of $\rho$. The partial transposed
matrix of $\rho$ is given by \be \rho^{T_1}=\sum_{l,j,k=1}^n
x_{l}^{j}\,\overline{x}_{l}^{k}|k\rangle\langle
j|\otimes|j+l-1\rangle\langle k+l-1|, \ee where $T_1$ stands for
partial transpose with respect to the first Hilbert space. That
$\rho$ is PPT means that $\rho^{T_1}\geq 0$. Namely, for any vector
$|Y\rangle=\sum_{r,s=1}^{n}y_r^{s+r-1}|r\rangle\otimes|s\rangle$ in
$H\otimes H$, we obtain $\langle {Y}|\rho^{T_1}|Y\rangle\geq 0$.
Here and later, we use $s+r-1$ to represent $s+r-1\ mod\ n$, $mod$
denoted modulo arithmetic. We have
$$
\begin{array}{rcl}
\langle {Y}|\rho^{T_1}|Y\rangle &=&\displaystyle\langle
{Y}|\sum_{l,j,k=1}^n x_{l}^{j}\,\overline{x}_{l}^{k}|k\rangle\langle
j|\otimes|j+l-1\rangle\langle k+l-1|\sum_{r,s=1}^{n}y_r^{s+r-1}|r\rangle\otimes|s\rangle\\
&=&\displaystyle\langle{Y}|\sum_{l,j,k=1}^nx_{l}^{j}\,\overline{x}_{l}^{k}\,y_r^{s+r-1} \delta_{j}^{r}\,\delta_{s}^{(k+l-1)}|k\rangle\otimes|j+l-1\rangle\\
&=&\displaystyle\sum_{r^\prime,s^\prime=1}^{n}y_{r^{\prime}}^{s^\prime+r^\prime-1}|r^\prime
\rangle\otimes|s^\prime\rangle\sum_{l,j,k=1}^nx_{l}^{j}\,
\overline{x}_{l}^{k}\,y_j^{k+j+l-2}|k\rangle\otimes|j+l-1\rangle\\
&=&\displaystyle\sum_{l,j,k,r^\prime,s^\prime=1}^{n}x_{l}^{j}\,
\overline{x}_{l}^{k}\,y_j^{k+j+l-2}\,\overline{y}_{r^\prime}^{r^\prime+s^\prime-1}\,
\delta_{r^\prime}^k\,\delta_{s^\prime}^{j+l-1}\\
&=&\displaystyle\sum_{l,j,k=1}^{n}x_{l}^{j}\,
\overline{x}_{l}^{k}\,y_j^{k+j+l-2}\,\overline{y}_{k}^{k+j+l-2}\geq
0.
\end{array}
$$
Because of the independence of the variables $y_r^\alpha$, the above
inequality is equivalent to the following inequities:
$$\begin{array}{rcl} <y^{n-1}|A_1|y^{n-1}>&\equiv&
\displaystyle\sum_{l,j,k=1, l+j+k=1\, mod \ n}^{n}x_{l}^{j}\,
\overline{x}_{l}^{k}\,y_j^{n-1}\,\overline{y}_{k}^{n-1}\geq 0,\\[6mm]
<y^{n}|A_2|y^n>&\equiv& \displaystyle\sum_{l,j,k=1, l+j+k=2\, mod \
n}^{n}x_{l}^{j}\,
\overline{x}_{l}^{k}\,y_j^{n}\,\overline{y}_{k}^{n}\geq 0,\\
\cdots&&\\
<y^{n-3}|A_{n-3}|y^{n-1}>&\equiv& \displaystyle\sum_{l,j,k=1,
l+j+k=n-1\, mod \ n}^{n}x_{l}^{j}\,
\overline{x}_{l}^{k}\,y_j^{n-3}\,\overline{y}_{k}^{n-3}\geq 0,\\
<y^{n-2}|A_n|y^{n-2}>&\equiv& \displaystyle\sum_{l,j,k=1, l+j+k=0\,
mod \ n}^{n}x_{l}^{j}\,
\overline{x}_{l}^{k}\,y_j^{n-2}\,\overline{y}_{k}^{n-2}\geq 0,
\end{array}$$
where $|y^i>=(y^i_1,y^i_2,...,y^i_n)^t$ ($t$ stands for transpose),
$i=1,2,...,n$. $A_1,A_1,\cdots,A_{n}$ are non-negative, hermitian
matrices, with the entries of $A_m$ given by $x_{l}^{j}
\overline{x}_{l}^{k}$ for $l+j+k=m\, mod\, (n)$. For fixed $m$,
$l=[m-(j+k)]\, mod\, n$ only depends on $j,k$.

For example, when $n=5$, one has
\begin{equation}
A_1=\left(
  \begin{array}{ccccc}
    x_4^1\overline{x}_4^1& x_3^1\overline{x}_3^2 & x_2^1\overline{x}_2^3 & x_1^1\overline{x}_1^4 & x_5^1\overline{x}_5^5 \\
    x_3^2\overline{x}_3^1& x_2^2\overline{x}_2^2 & x_1^2\overline{x}_1^3 & x_5^2\overline{x}_5^4 & x_4^2\overline{x}_4^5 \\
    x_2^3\overline{x}_2^1& x_1^3\overline{x}_1^2 & x_5^3\overline{x}_5^3 & x_4^3\overline{x}_4^4 & x_3^3\overline{x}_3^5 \\
    x_1^4\overline{x}_1^1& x_5^4\overline{x}_5^2 & x_4^4\overline{x}_4^3 & x_3^4\overline{x}_3^4 & x_2^4\overline{x}_2^5 \\
    x_5^5\overline{x}_5^1& x_4^5\overline{x}_4^2 & x_3^5\overline{x}_3^3 & x_2^5\overline{x}_2^4 & x_1^5\overline{x}_1^5
  \end{array}
\right), \cdots.
\end{equation}

Due to the non-negativity of the
matrices $A_1$, ..., $A_{n}$, all the principal minors of $A_m$, $m=1,2,...,n$,
are non-negative. We have

\noindent {\bf Theorem 1:} The entries of the matrices $A_m,\
\forall \ m=1, 2,\cdots, n$ satisfy the following quadratic
relations,
\begin{equation}\label{cond 1}
x^{ip} x^{jq}=x^{iq}x^{jp},
\end{equation}
where $x^{ip}=x^{i}_{l(i,p)}\overline{x}^{p}_{l(i,p)}$, $l(i,p)=(m\,
mod\, n)-(i+p)\equiv m-(i+p)$, the other marks have the same meaning.

\noindent {\bf Proof:} First, we consider order two principal minors
$\{(i,j),(i,j)\}$ of the matrix $A_m$. From the non-negativity of
$A_m$, we get that $x^{ii}x^{jj}\geq x^{ij}x^{ji}$. The inequality
is in fact an equality. Because if for some $m$, $x^{ii}x^{jj}>
x^{ij}x^{ji}$, then
$\prod_{m=1}^{n}x^{ii}x^{jj}>\prod_{m=1}^{n}x^{ij}x^{ji}$. On the
other hand, from straightforward calculation, we have
$\prod_{m=1}^{n}x^{ii}x^{jj}=\prod_{m=1}^{n}x^{ij}x^{ji}$ for fixed
$i,j$. Therefore, for any $m$ we have \be\label{condition 1}
x^{ii}x^{jj}= x^{ij}x^{ji}. \ee

Second, from the non-negativity of the order three principal minors
$\{(i,j,k),(i,j,k)\}$ of the matrix $A_m$, we have $0\leq
x^{ij}x^{jk}x^{ki}-x^{ii}x^{jk}x^{kj}+x^{ik}x^{ji}x^{kj}-x^{ik}x^{jj}x^{ki}
=x^{ij}x^{jk}x^{ki}-2x^{ii}x^{jj}x^{kk}+x^{ik}x^{ji}x^{kj}
=2Re(x^{ij}x^{jk}x^{ki})-2x^{ii}x^{jj}x^{kk}\leq
2(|x^{ij}||x^{jk}||x^{ki}|-x^{ii}x^{jj}x^{kk})=
2(\sqrt{x^{ii}x^{jj}}\sqrt{x^{jj}x^{kk}}\sqrt{x^{kk}x^{ii}}-x^{ii}x^{jj}x^{kk})=0$,
where we have used the condition (\ref{condition 1}) in the first
and the third equations. Therefore, we get the following relations:
\be\label{condition 2} x^{ij}x^{jk}x^{ki}=x^{ii}x^{jj}x^{kk}. \ee
For a nonzero $3\times 3$ hermitian matrix, if its order two and
three principal minors are all $0$, then it has only one eigenvalue,
and all of its order two minors are $0$. Therefore, we have
\be\label{condition 3} x^{ij}x^{jk}=x^{ik}x^{jj}. \ee

Third, combining the non-negativity of the order four principal minors
$\{(i,j,k,l),(i,j,k,l)\}$ of the matrix $A_m$ with
(\ref{condition 1}) and (\ref{condition 2}), we have
$$
6x^{ii}x^{jj}x^{kk}x^{ll}-2Re( x^{il}x^{lk}x^{kj}x^{ji}+x^{jl}x^{lk}x^{ki}x^{ij}+ x^{jl}x^{li}x^{ik}x^{kj})\geq 0.
$$
Using relations (\ref{condition 3}), we have
$x^{ii}x^{jj}x^{kk}x^{ll}-Re
(x^{il}x^{lk}x^{kj}x^{ji})=x^{ii}x^{jj}x^{kk}x^{ll}-Re(
x^{ii}x^{jj}x^{kk}x^{ll})=x^{ii}x^{jj}x^{kk}x^{ll}-
x^{ii}x^{jj}x^{kk}x^{ll}=0$. Hence all order four principal minors
are all $0$ and $x^{il}x^{lk}x^{kj}x^{ji}=x^{ii}x^{jj}x^{kk}x^{ll}$.
Since order two, three and four principal minors are equivalent to
$0$, therefore, the nonzero $4\times 4$ hermitian matrix (order four
principal minors) has only one eigenvalue, then all of its order two
minors are $0$. Furthermore, all order two minors are included in
one order four principal minors. Therefore the entries of $A_m$
satisfy the relations (\ref{cond 1}). \qed

From the condition (\ref{condition 1}),
$x_{m-2j}^{j}\,\overline{x}_{m-2j}^{j}\,x_{m-2k}^{k}\,\overline{x}_{m-2k}^{k}=
x_{m-(j+k)}^{j}\,\overline{x}_{m-(j+k)}^{k}\,x_{m-(j+k)}^{k}\,\overline{x}_{m-(j+k)}^{j}$,
we have the following relations: \be\label{condition 4}
x_{m-(j+k)}^j\,x_{m-(j+k)}^k\,e^{i\theta_m^{jk}}=x_{m-2j}^j\,x_{m-2k}^k
\ee or \be\label{condition 5}
x_{m}^j\,x_{m}^k\,e^{i\theta_m^{jk}}=x_{m-(j-k)}^j\,x_{m+(j-k)}^k,
\ee where $0\leq\theta_m^{jk}\leq2\pi$.

\noindent {\bf Theorem 2:}  The number of independent $\theta_m^{jk}$ is at most $n-1$.

\noindent {\bf proof:} First, from (\ref{condition 3}),we have
$x^{ij}x^{jk}=x^{ik}x^{jj}$, i.e,
$x_{m-(i+j)}^i\overline{x}_{m-(i+j)}^jx_{m-(j+k)}^j
\overline{x}_{m-(j+k)}^k=x_{m-(i+k)}^i\overline{x}_{m-(i+k)}^kx_{m-2j}^j\overline{x}_{m-2j}^j$.
Let $j-i=k-j$. We obtain
$$
x_{m-(i+j)}^i\overline{x}_{m-(i+j)}^jx_{m-(j+k)}^j\overline{x}_{m-(j+k)}^k=x_{m-2j}^i
\overline{x}_{m-2j}^kx_{m-2j}^j\overline{x}_{m-2j}^j,
$$
which gives rise to
$x_{m-i+j}^i\overline{x}_{m-i+j}^jx_{m+j-k}^j\overline{x}_{m+j-k}^k=x_{m}^i\overline{x}_{m}^kx_{m}^j\overline{x}_{m}^j$.
Namely,
$x_{m+s}^i\overline{x}_{m+s}^jx_{m-s}^j\overline{x}_{m-s}^k=x_{m}^i\overline{x}_{m}^kx_{m}^j\overline{x}_{m}^j$.
However from (\ref{condition 4}) we have
$x_{m+s}^ix_{m-s}^j=x_{m}^ix_{m}^je^{i\theta_m^{ij}}$ and
$\overline{x}_{m+s}^k\overline{x}_{m-s}^j=\overline{x}_{m}^k\overline{x}_{m}^je^{i(-\theta_m^{kj})}$.
Therefore, \be\label{condition 10}
\theta_m^{ij}=\theta_m^{kj}~~~~~if~j-i=k-j, \ee and $\theta_m^{ij}$
depends on the difference of $i$ and $j$.

Set $s=|i-j|$. In the following, we denote $\theta_m^{ij}$ as $\theta_m^{s}$. In
particular, we denote $\theta_m^1$ as $\theta_m$.
There are $[\frac{n}{2}]$ angles $\{\theta_m^s\}$ for given $m$, with
$s=1,2,\cdots,[\frac{n}{2}]$, $[x]$ denoting the integer that is less or equal to $x$.

Second, from (\ref{condition 5}) and (\ref{condition 10}), for any
integer $j\leq[\frac{n}{2}]$ and given $m$, we can get the following
equation:
$\prod_{l=0}^sx_{m+l}^jx_{m+l}^{j+1}e^{i\sum_{l=0}^s\theta_{m+l}}=\prod_{l=0}^sx_{m+l+1}^jx_{m+l-1}^{j+1}$.
That is \be\label{condition
11}x_m^jx_{m+s}^{j+1}e^{i\sum_{l=0}^s\theta_{m+l}}=x_{m+s+1}^jx_{m-1}^{j+1}.
\ee Following (\ref{condition 11}), we can get $s$ equations:
$x_m^jx_{m+s-1}^{j+1}e^{i\sum_{l=0}^{s-1}\theta_{m+l}}=x_{m+s}^jx_{m-1}^{j+1}$,
$x_{m-1}^{j+1}x_{(m+s-1)-1}^{j+2}e^{i\sum_{l=0}^{s-1}\theta_{m-1+l}}=x_{m+s-1}^{j+1}x_{m-1-1}^{j+2}$,
$\cdots$,
$x_{m-(s-1)}^{j+s-1}x_{m}^{j+s}e^{i\sum_{l=0}^{s-1}\theta_{m-(s-1)+l}}=x_{m+s-(s-1)}^{j+s-1}x_{m-1-(s-1)}^{j+s}$.
Multiplying these equations together, we get
$$
x_m^jx_m^{j+s}e^{i(s\theta_m+(s-1)(\theta_{m-1}+\theta_{m+1})+\cdots+(\theta_{m+s-1}+\theta_{m-s+1}))}=
x_{m+s}^jx_{m-1-(s-1)}^{j+s}=x_m^jx_m^{j+s}e^{i(\theta_m^s)},
$$
i.e. any $\theta_m^{s}$, $s\geq 2$, $m= 1,2,\cdots,n$ can be expressed according to the
angles $\theta_m$, $m=1,2,\cdots,n$.

Furthermore, for fixed $j,k$ or $s$, we have
$\prod_{m=0}^{n-1}x_{m-2j}^jx_{m-2k}^k=\prod_{m=0}^{n-1}x_{m-(j+k)}^jx_{m-(j+k)}^k
e^{i\Sigma_{m=0}^{n-1}\theta_m^{jk}}$. On the other hand, by
direct computation, we have
$\prod_{m=0}^{n-1}x_{m-2j}^jx_{m-2k}^k=\prod_{m=0}^{n-1}x_{m-(j+k)}^jx_{m-(j+k)}^k$.
Hence $\Sigma_{m=0}^{n-1}\theta_m^{jk}=0$, or
$\sum_{m=0}^{n-1}\theta_m^{s}=0$, $s=1,2,\cdots,[\frac{n}{2}]$.
Therefore, there are in fact only $n-1$ independent angles $\theta_m$.
\qed

For example, using (\ref{condition 5}) we have
$x_1^1x_1^2e^{i\theta_1}=x_2^1x_n^2$,
$x_2^1x_2^2e^{i\theta_2}=x_3^1x_1^2$. Hence
$x_1^1x_2^2e^{i(\theta_1+\theta_2)}=x_3^1x_n^2$. From
$x_1^2x_1^3e^{i\theta_1}=x_2^2x_n^3$ and
$x_n^2x_n^3e^{i\theta_n}=x_1^2x_{n-1}^3$, we
get $x_1^3x_n^2e^{i(\theta_1+\theta_n)}=x_2^2x_{n-1}^3$,
which give rise to
$x_1^1x_1^3e^{i(\theta_1^{13})}=x_1^1x_1^3e^{i(\theta_1^{12}+\theta_2^{12}
+\theta_1^{23}+\theta_n^{23})}=x_3^1x_{n-1}^3$,
i,e.
\begin{equation}\label{e1}
\theta_1^{13}=\theta_1^{2}=(2\theta_1+\theta_2+\theta_n).
\end{equation}

We are now ready to construct pure separable state decompositions of
$\rho$ when $\rho$ is PPT. Let $U$ be a unitary transformation, with
its entries given by
$u_{kl}=(\frac{1}{\sqrt{n}}e^{i((k-1)(l-1)\omega+\delta_k)})$, where
$\delta_k$ $k=1,2,\cdots,n$ is an angle, $\omega$ is the n-th unit
root, $\omega^n=1$. Then $\rho=\sum_{l=1}^n|X_l\rangle\langle
X_l|=\sum_{l=1}^n|Z_l\rangle\langle Z_l|$, where
\begin{equation}\label{sd}
|Z_l\rangle=\sum_{k=1}^{n}u_{kl}|X_l\rangle=\sum_{i,j=1}^{n}b_{rs}^l|rs\rangle.
\end{equation}
Denoting $B_l=(b_{rs}^l)$, one has
$$
B_l=(b_{rs}^l)=((e^{i((s-r)(l-1)\omega+\delta_{s-r+1})}x_{s-r+1}^{r})_{rs}).
$$
For example, when $n=5$, one has
\begin{equation}
B_l=\left(
  \begin{array}{ccccc}
    u_{1l}x_1^1 & u_{2l}x_2^1 & u_{3l}x_3^1 & u_{4l}x_4^1 & u_{5l}x_5^1 \\
    u_{5l}x_5^2 & u_{1l}x_1^2 & u_{2l}x_2^2 & u_{3l}x_3^2 & u_{4l}x_4^2 \\
    u_{4l}x_4^3 & u_{5l}x_5^3 & u_{1l}x_1^3 & u_{2l}x_2^3 & u_{3l}x_3^3 \\
    u_{3l}x_3^4 & u_{4l}x_4^4 & u_{5l}x_5^4 & u_{1l}x_1^4 & u_{2l}x_2^4 \\
    u_{2l}x_2^5 & u_{3l}x_3^5 & u_{4l}x_4^5 & u_{5l}x_5^5 & u_{1l}x_1^5
  \end{array}
\right).
\end{equation}

\noindent {\bf Theorem 3:} There exist $\delta_k$ such that every
order two minors $\{(m,k), (\alpha,\beta)\}$ in $B_l$ is zero, and so that
$\rho=\sum_{l=1}^n|Z_l\rangle\langle Z_l|$ is a pure separable state decomposition
for $\rho$ that is PPT.

\noindent {\bf Proof} \ That any order two minors $\{(m,k),
(\alpha,\beta)\}$ of $B_l$ are zero implies: \be\label{condition 6}
e^{i(\delta_{\alpha-m+1}+\delta_{\beta-k+1}-\delta_{\alpha-k+1}
-\delta_{\beta-m+1})}x_{\alpha-m+1}^mx_{\beta-k+1}^k=
x_{\beta-m+1}^mx_{\alpha-k+1}^k. \ee Namely, any order two minors
$\{(m,m+1), (\alpha,\alpha+1)\}$ should be zero, \be\label{condition
7}e^{i(2\delta_{\alpha-m+1}-\delta_{\alpha-m+2}-\delta_{\alpha-m})}
x_{\alpha-m+1}^mx_{\alpha-m+1}^{m+1}=x_{\alpha-m+2}^mx_{\alpha-m}^{m+1}.
\ee From the PPT conditions, we have:
$x_m^jx_m^{j+1}e^{i\theta_m}=x_{m+1}^jx_{m-1}^{j+1}$. Applying
Theorem 2, we have
$x_m^jx_m^{j+1}e^{i(2\delta_m-\delta_{m+1}-\delta_{m-1})}=x_{m+1}^jx_{m-1}^{j+1}=x_m^jx_m^{j+1}e^{i\theta_m}$.
Therefore \be\label{condition 8}
2\delta_i-\delta_{i+1}-\delta_{i-1}=\theta_i, i=1,2,\cdots,n. \ee
Eq. (\ref{condition 8}) has always solutions for $\delta_{i}$ with
the relationship $\sum_{i=1}^n\theta_i=0$. As every order two minors
$\{(m,k), (\alpha,\beta)\}$ of $B_l$ is zero, the rank of $B_l$ is
one. Therefore, $|Z_l\rangle$ is separable. \qed

In fact the solutions of Eq. (\ref{condition 8}) are not unique. By
calculating, we know that there is a free variable of the parameters
$\delta_m$, $m=1, 2,\cdots,n$, therefore exist many different
separable pure state decompositions for such $\rho$.

We have investigated a class of $n\otimes n$ bipartite mixed states
for which the PPT criterion is both sufficient and necessary for
separability. The PPT conditions for these states are derived. We
have presented a general approach to find the separable pure state
decompositions of this class, and the separable pure state
decompositions have been explicitly constructed.

\textbf{Acknowledgments}

Huang gratefully acknowledges the grant support from the Natural
Sciences Foundation for Hainan Province 111006. Jing is grateful for
the support of Simons Foundation grant 198129 and NSFC grant
11271138 during the work. This work is also supported by the NSFC
under number 11105226. We would like to acknowledge helpful
discussions with Bobo Hua.

\end{document}